\documentclass[11pt]{article}
\usepackage{amsmath,amssymb}
\newtheorem{thm}{Theorem}[section]

\newtheorem{lem}[thm]{Lemma}
\newtheorem{pr}[thm]{Proposition}
\newtheorem{definition}[thm]{Definition}

\newtheorem{example}[thm]{Example}

\newtheorem{remark}[thm]{Remark}

\newcommand{\F}{\mathbb{F}}

\newcommand{\Section}[1]{\section{#1}\setcounter{equation}{0}}

\newcommand{\openbox}{\leavevmode
  \hbox to.77778em{%
    \hfil\vrule
  \vbox to.675em{\hrule width.6em\vfil\hrule}%
  \vrule\hfil}}
\newcommand{\proofname}{Proof}

\newenvironment{proof}[1][\proofname]{\par\normalfont
  \trivlist\item[\hskip\labelsep\itshape #1:]\ignorespaces
  }{\hspace*{1cm}\hspace*{\fill}\openbox \medskip\endtrivlist}

\title{
  Analysis of a Key Distribution Scheme in Secure Multicasting
  \footnote{Partially supported by SNF Grant No. 121874. To appear in
    {\it Journal of Mathematical Cryptology}, Vol. 6, 1 (2012), pp. 69--80.}
}%
\date{}%
\author{G\'erard Maze\\
{\small {\em e-mail:\/} gmaze@math.uzh.ch \vspace{-1mm} }\\
{\small Mathematics Institute\vspace{-1mm}}\\
{\small University of Z\"urich\vspace{-1mm}}\\
{\small Winterthurerstr 190, CH-8057 Z\"urich, Switzerland }
\vspace{3mm} }

\begin{document}\maketitle
\thispagestyle{empty}
\begin{abstract}
  This  article presents an  analysis of  the secure  key broadcasting
  scheme proposed by  Wu, Ruan, Lai and Tseng  \cite{wu}. The study of
  the parameters of the system is based on a connection with a special
  type  of symmetric  equations over  finite fields.   We  present two
  different attacks  against the  system, whose efficiency  depends on
  the choice of the parameters.  In particular, a time-memory tradeoff
  attack is  described, effective  when a parameter  of the  scheme is
  chosen without care. In such a situation, more than one third of the
  cases can be broken with a time and space complexity in the range of
  the square root of the complexity of the best attack suggested by Wu
  et al.  against their system.  This  leads to a feasible attack in a
  realistic scenario.
\end{abstract}
\vspace{3mm}
\noindent{\bf Key Words:} Finite fields, time-memory tradeoff attack, 
system of power equations\\
\noindent{\bf Subject Classification:} 11T55, 94A60, 68P30

%%%%%%%%%%%%%%%%%%%%%%%%%%%%%%%%%%%%%%%%%%%%%%%%%%%%%%%%%%%%%%%%%%%
\Section{Introduction}

The  goal  of  this  article  is  to present  an  analysis  of  a  key
distribution scheme taking place  in a multicasting system. The system
has been developed by Wu, Ruan, Lai and Tseng, see \cite{wu}, in order
to propose a new solution to the problem of transmitting securely keys
in the context of multicasting. In such a context, the security of the
transmission  must be  coupled with  the imperative  of being  able to
manage groups of users sharing  the same key where typically one wants
to deal with users leaving a  group after some time, new users joining
different groups,  etc...  The solution  of Wu et  al.  is based  on a
particular finite  field construction and  its security relies  on the
computational difficulty  of a problem  that appears to have  not been
studied rigorously up to now.\\

The  problem, presented  in more  details in  Section  \ref{Wu} below,
takes  place in  a finite  prime field  $\F_p$ where  a  $n$-th degree
polynomial $f$ is  given and consists in finding $k  \in \F_p$ so that
$f(x)-k$ splits  into linear factors  in $\F_p$, provided that  such a
$k$ exists. We  will see that this problem  is directly connected with
the so-called {\it systems of power equations} \cite{mc1,mc2}. Indeed,
the problem is equivalent to  solving an inhomogeneous system of $n-1$
power equations in $n$ variables with degrees running from 1 to $n-1$.
This type  of equations  with symmetries are  known to  be generically
hard to  solve computationally, see  e.g.  \cite{bettale,faugere}, and
they often appear  as test case when evaluating  algorithms whose goal
is  to  find  solutions  of  systems  of  polynomial  equations.   For
instance, at  the time of writing, these  problems are computationally
intractable as  soon as the  degree of the  system is as large  as 30,
even in a finite field  with moderate size.  In the current situation,
the degree  of the system  can potentially be  a few thousand  and the
finite  field size  should  be  chosen larger  than  $2^{80}$.  It  is
however worth  noticing that the  special form of the  equations $S_n$
described below might turn out to  be in fact easily solvable, but the
author  of  the  article  is  unaware  of  any  algorithm  capable  of
performing this task efficiently.\\

Even though the connection with systems of cyclic power equations does
not lead to  a feasible computational solution of  the initial problem
underlying the  system of Wu et al.,  this link will allow  us to shed
light on the expected number of solutions of the initial problem. This
will be  explained in Section \ref{cy}. Since  Gr\"obner bases methods
as  well  as  different  linearization  techniques do  not  appear  to
threaten the security of the system in the generic case, we will focus
in Section  \ref{At} on  the case  where the order  $p$ of  the finite
field  has been  chosen without  care. Based  on this  assumption, two
different  attacks will  be  presented. In  particular, a  time-memory
tradeoff  attack  against the  system  will  be  developed whose  time
complexity $T$ and memory complexity $M$ satisfy $TM=O(p \ln^3 p)$ and
are both in the order of the square root of $p$ in more than one third
of the  cases. We  would like to  point out  that in such  a realistic
situation the time-memory tradeoff  attack can be potentially realized
on  a system where  the parameters  have been  chosen as  described in
\cite{wu}.\\

All the  computations and equalities  in this article should  be clear
from the  context. The natural logarithm  and the logarithm  in base 2
are denoted  by $\ln$ and  $\log_2$ respectively.  We will  follow the
standard asymptotic notations, as in e.g.  \cite{gathen}, such as $o$,
$O$ and $\ll$. We will write  $f(n) \geqslant g(n) (1+o(1)) $ when $f$
and $g$ satisfy $\liminf_{n \rightarrow \infty } (f(n)/g(n)-1) > 0$.

%%%%%%%%%%%%%%%%%%%%%%%%%%%%%%%%%%%%%%%%%%%%%%%%%%%%%%%%%%%%%%%%%%%
\Section{The Key Broadcasting Scheme of Wu et al.}\label{Wu}

Let  us now  present the  technical  details of  the key  distribution
scheme in secure multicasting of Wu, Ruan, Lai and Tseng. We refer the
reader to the original paper \cite{wu} for a more detailed description
of the broadcasting  setting and on the argumentation  of the benefits
of the system. The ground  parameters of the multicasting system are a
large finite  prime field  $\F_p$ and a  family $\mathcal{H}$  of hash
functions with  values in $\F_p$. Each  user of the  system receives a
private key $a  \in \F_p $ that  is fixed for a given  time period and
that is known to the key management authority. When the key management
authority wants to  broadcast a key $k$ to  $n$ distinguished users of
the  system with  private  keys $a_1,\ldots,a_n$,  he  selects a  hash
function  $h \in  \mathcal{H}$  and expands  the  monic $n$-th  degree
polynomial $f$ in $\F_p[x]$ as follows:
\begin{equation}\label{poly}
f(x) = \prod_{i=1}^n (x-h(a_i)) + k = x^n + \sum_{j=0}^{n-1} b_{n-j} x^j.
\end{equation}
The management authority  sends to the $n$ users  the $n$ coefficients
$b_j$ together with  the hash function $h$. Since  the polynomial $f$,
the so-called  ``secure filter'' in \cite{wu},  satisfies $f(h(a_i)) =
k$ for all $i=1,\ldots,n$, the $n$ distinguished users can compute the
key $k$. The  system is secure in the sense  that an unauthorized user
who wants  to have  access to  $k$ faces the  problem to  recover this
field element from the broadcasted parameters $b_0,b_1,\ldots,b_{n-1}$
and $h$. The  designers of the system state  in \cite[Section 3.3]{wu}
that $k$ can only be obtained  from the constant term $b_0$ since $b_0
=  k+\prod_{i=1}^n h(a_i)$  and not  knowing the  $h(a_i)$'s  makes it
infeasible because the finite field size $p$ is too large.\\

The  distribution  of  the  $n$  field  elements  $b_i$  represents  a
transmission  of  $n\log_2(p)$ bits.   The  distribution  of the  hash
function is  not explained in the original  setting \cite{wu}, however
in order to balance the security between the choice of the key $k$ and
the function  $h$, the number of  possible hash functions  should be at
least as  large as $p$.  For instance,  it would be possible  to fix a
cryptographic  hash function  $h$, and  define $\mathcal{H}  =  \{ h_c
\}_{c \in \F_p}$ where $ h_c(x) = h(h(x)+c)$. In doing so, any element
of $\cal{H}$  is described  with a field  element.  We  will therefore
assume  that the  key  distribution requires  $O(n\log_2(p))$ bits  of
transmission.   This is  however  not a  limiting  requirement in  our
analysis. 
When a fixed hash function $h$ is used for each broadcasting, the
system is not immune against attacks during different phases of the
scheme, as described in \cite{zhu}. However, when the hash function is
different for each transmission, as suggested as vulnerabilty fix in
\cite{zhu}, the system becomes exactly the one described above.
We would like to point out that it is in the interest of the
designer to  select the size of  $p$ in order to  balance the security
and the  transmission cost.  In a multicasting  environment, the value
$n$ can potentially be quite big  (up to a few millions), leading to a
choice of  the size  of $p$  as small as  the security  concerns would
allow. With this in mind, we  will naturally assume in the sequel that
$n<p$. \\

The brute force attack suggested  by the authors relies on testing the
$p$ possible keys $k \in \F_p$. This exhaustive search can potentially
be directly operated on the system  the key is supposed to enable, but
it  is  also  possible  to  run the  following  algebraic  test.   The
polynomial  $f$ and the  key $k$  are such  that $f(x)-k$  splits into
linear  factors over  $\F_p$.  This  means that  $f(x)-k$  divides the
product of all  linear monic polynomials, which is  $x^p-x$, see e.g.,
\cite{lidle}. This is equivalent to write that
\begin{equation}\label{eq}
x^p-x = 0 \mod (f(x)-k).
\end{equation}
Testing the  previous equality can  be done in  $O(\log_2(p))$ modular
polynomial operations,  using repeated square  and multiply techniques
in the ring $\F_p[x]/(f(x)-k)$,  see e.g. \cite{menezes}. Any $k$ that
fulfills the previous equation is  a candidate. The expected number of
candidates is analyzed  in the next section and turns  out to be small
as soon  as $n =  \frac{\ln p }{\ln\ln  p} (1+o(1))$. This leads  to a
brute  force attack  with  time complexity  $O(p\log_2(p))$ and  space
complexity  $O(n\log_2(p))$ when  $n$  is large  enough.  A  realistic
situation could  be the  following.  The finite  field is  selected to
have $p \cong  2^{75}$ elements, so that the brute  force attack has a
complexity  of more  than $2^{80}$  modular polynomial  operations. As
soon  as $n  \geqslant  15$, only  a  few $k  \in  \F_p$ will  satisfy
Eq.~(\ref{eq}).  With $n=100000$ users (a  factor of 40 less than some
currently  used pay-TV systems  \cite{wiki}), the  multicasting system
would need to broadcast almost 1 megabyte of information.

%%%%%%%%%%%%%%%%%%%%%%%%%%%%%%%%%%%%%%%%%%%%%%%%%%%%%%%%%%%%%%%%%%%
\Section{Connection with Systems of Power Equations}\label{cy}

Our first  goal is  to find  an estimation of  the number  of possible
candidates $k \in \F_p$ satisfying Eq.~(\ref{eq}) and to determine how
difficult it is to  compute one of these.  In order to  do so, we will
make use of  a special type of polynomial  equations over $\F_p$.  Let
us  consider $S_n=S_n(s_1,\ldots,s_{n-1})$,  the  following system  of
$n-1$ consecutive power equations in $n$ variables:
\begin{eqnarray*}
x_1 \; \; + \; \;x_2 \; \;+\; \ldots \; + \; \;x_n \;& = & s_1\\
x^2_1\; \; +\; \; x^2_2\; \; +\; \ldots\; + \; \; x^2_n \;& = & s_2\\
x^3_1\; \; +\; \; x^3_2\; \; +\; \ldots\; +\; \; x^3_n \;& = & s_3\\
\vdots \hspace{1.6cm} & = & \; \vdots\\
x^{n-1}_1 + x^{n-1}_2 + \ldots + x^{n-1}_n & = & s_{n-1}\\
\end{eqnarray*}
Notice  that  if  one more  power  equation  of  degree $n$  would  be
available, then  the system would  be solvable in  expected polynomial
time, see e.g.   \cite{mc1,mc2} and \cite{lidle} for the  use of it in
decoding BCH codes. The above  system is non-trivial because this last
equation is missing. Recall that the coefficients of the polynomial
\begin{equation}\label{tt}
\prod_{j=1}^n (x-x_j) = x^n + \sum_{j=0}^{n-1} e_{n-j} x^j
\end{equation}
are explicitly related to the sum of the powers of its roots $x_j$ via
Newton's  identities,   that  have   the  following  form,   see  e.g.
\cite{mead},
\begin{equation}\label{newton}
e_j = F_j(s_1,\ldots,s_{j-1})- (-1)^j \frac{s_j}{j}
\end{equation}
for  some  specific  algebraically  independent polynomials  $F_j  \in
\F_p[y_1,\ldots,y_{j-1}], \;  j >  0$. For instance,  $e_1 =  s_1$ and
$e_2 = \frac{s_1^2}{2}-\frac{s_2}{2}$. The special triangular shape of
the  equations (see  e.g.   \cite{mead}), i.e.,  the  fact that  $F_j$
depends  on  $s_1,\ldots,  s_{j-1}$  only, together  with  the  affine
dependence between $e_j$ and $s_j$ has several implications.

First,  one  can recursively  compute  $s_j$  for $j=0,\ldots,n-1$  in
polynomial time  as soon as  the $e_j$ are given  for $j=0,\ldots,n-1$
(note that  since we assumed  $n<p$, the division  by $j$ in  the last
term  of (\ref{newton}) is  never a  problem).  Therefore  solving the
initial problem  (\ref{eq}) with unknown $k$ is  equivalent to solving
the system  $S_n$ with $b_j=e_j$ for  $j=0,\ldots,n-1$ since computing
any $x_i=h(a_i)$ is essentially equivalent to computing $k$ (factoring
splitting  polynomials in $\F_p$  can be  done in  expected polynomial
time).   This  gives some  confidence  in  the  general difficulty  of
breaking  the multicasting  scheme, since  solving $S_n$  for randomly
chosen  $s_1,\ldots,  s_{n-1}$  seems  to  be  a  difficult  task,  as
explained in the introduction.

Second, the number  of solutions of $S_n$ is related  to the number of
possible $k$ such that (\ref{eq})  holds. If we consider two solutions
of  $S_n$ to  be the  same if  one  is obtained  from the  other by  a
permutation of its  components, then there is a  bijection between the
set  of solutions  of $S_n$  and the  set of  possible  $k$ satisfying
(\ref{eq}). Indeed, if $(x_1,\ldots,x_n)$  is a solution of $S_n$ then
$k =  f(x_0)-\prod_{i=1}^{n-1} x_i$ satisfies (\ref{eq}),  and any $k$
satisfying (\ref{eq}) gives a completely splitting polynomial $f(x)-k$
with a unique set of roots,  up to permutations.  If $\Omega_n$ is the
set of unordered $n$-tuples of  elements of $\F_p$, then a solution of
$S_n$ is an element of $\Omega_n$ and $|\Omega_n| = \binom{p+n-1}{n}$.

Finally,  let  us  focus  on  the  expected  number  of  possible  $k$
satisfying (\ref{eq}), when the  coefficients of the polynomial $f(x)$
are independently  and uniformly distributed at random  in $\F_p$. The
triangular shape and the  affine dependence described above imply that
the $s_i$ are independently and uniformly distributed in $\F_p$ if and
only if the same is true for  the $e_i$. This comes from the fact that
if  $X$ and  $Y$  are  independent random  variables,  with $Y$  being
uniformly distributed, then $X+Y$ is uniformly distributed. Therefore,
when  the  $n-1$  coefficients  of  strictly positive  degree  of  the
polynomial $f(x)$ are chosen  independently and uniformly at random in
$\F_p$, the expected  number of $k$ satisfying (\ref{eq})  is equal to
the  expected number  $N$ of  solutions  of $S_n(s_1,\ldots,s_{n-1})$,
when $s_1,\ldots,s_{n-1}$ are  independently and uniformly distributed
at random in $\F_p$. For $a \in \Omega_n$, let us write $1_{S_n(a)=0}$
for the  indicator function  of the set  $\{a \in  \Omega_n \; |  \; a
\mbox{ is a  solution of } S_n\}$.  The number $N$  can be computed as
follows:
\begin{eqnarray*}
N & = & \frac{1}{p^{n-1}} \sum_{s \in \F_p^{n-1}} |\{a \in \Omega_n \; | \; a
\mbox{ is a solution of } S_n(s) \}|\\
& = & \frac{1}{p^{n-1}} \sum_{s \in  \F_p^{n-1}} \sum_{a \in \Omega_n}
1_{S_n(s)(a)=0}\\
& = & \frac{1}{p^{n-1}} \sum_{a \in \Omega_n}  \sum_{s \in  \F_p^{n-1}}
1_{S_n(s)(a)=0}.
\end{eqnarray*}
Since  for  a  fixed  $a  \in  \Omega_n$ there  is  a  unique  $s  \in
\F_p^{n-1}$ such that $a$ is a solution of $S_n(s)$, we obtain that
\[
N = \frac{1}{p^{n-1}} \sum_{a \in \Omega_n} 1 =
\frac{\binom{p+n-1}{n}}{p^{n-1}}.
\]
Let us summarize the situation with the following lemma:
\begin{lem}\label{l1}
Let  $b_1,\ldots,b_{n-1}$ be  independently and  uniformly distributed
elements  in $\F_p$  and let  $f(x) =  x^n +  \sum_{j=1}^{n-1} b_{n-j}
x^j$. The expected number of  elements $k \in \F_p$ such that $f(x)-k$
splits     into     linear      factors     in     $\F_p$     is     $
\frac{\binom{p+n-1}{n}}{p^{n-1}}$.
\end{lem}
In  the   context  of  the   secure  key  broadcasting   scheme  under
consideration, the previous lemma can  be used, since in this case the
$b_j$'s  being   obtained  by  evaluating   algebraically  independent
polynomials at values of a  cryptographic hash function, it is natural
to  consider that  they  will behave  like  independent and  uniformly
distributed random variables over $\F_p$. Notice that
\begin{equation}\label{as}
\frac{\binom{p+n-1}{n}}{p^{n-1}} = \frac{p}{n!} \cdot \prod_{i=1}^{n-1}
\left(1+\frac{i}{p}\right) =  \frac{p}{n!} \cdot
\exp\left(\frac{n^2}{2p} + o\left(\frac{n^2}{p}\right)\right).
\end{equation}
This asymptotic expression invites us to separate two situations, when
$n  = O(p^{1/2})$ and  when $n$  is essentially  larger.  We  will not
address the  latter since it does  not fit any  plausible setting: the
prime  $p$ needs to  be very  large in  order to  give the  system its
security, and $n$ represents a  number of users, making the hypothesis
$n\gg p^{1/2}$ quite improbable. We  will therefore assume from now on
that  $n = O(p^{1/2})$  (even though  $n =  p^{1/2} (1+o(1))$  is also
quite  improbable).   In  this   situation,  the  expected  number  of
solutions essentially depends on the term $p/n!$.  We will make use of
the following technical lemma.
\begin{lem}\label{lambert}
If $n!=r$ then
\[
n = \frac{\ln(r/e)}{W(\frac{1}{e}\ln(r/e))} \cdot \left(1 +o(1) \right)
\]
where $W$ is the Lambert $W$ function that satisfies $W(t) \exp(W(t))
= t$ and 
\[
W(t) = \ln(t)\cdot \left(1 - \frac{\ln \ln t}{\ln t} +
\frac{\ln \ln t}{\ln^2 t} + o\left( \frac{\ln \ln t}{\ln^2 t} \right)
\right) = \ln(t)\cdot \left(1 +o(1) \right).
\]
\end{lem}
\begin{proof}
Since $\ln$  is increasing, we  have $\int_1^n \ln(x) \,  dx \leqslant
\sum_{i=1}^n \ln(i) \leqslant \int_1^n  \ln(1+x) \, dx$ and this leads
to     $e     (\frac{n}{e})^n     \leqslant     n!     \leqslant     e
(\frac{n+1}{e})^{n+1}$. By  continuity, there  exist $0<c<1$ with  $ e
(\frac{n+c}{e})^{n+c} =  r$. Thus $  (\frac{n+c}{e})^{\frac{n+c}{e}} =
(r/e)^{1/e}$. Solving  this equation for  $\frac{n+c}{e}$ is performed
with  the help  of the  Lambert $W$  function, defined  as  the unique
solution of the equation $W(t) \exp(W(t)) = t$ for $t\geqslant 0$, see
\cite{lambert}. In fact if $x^x=y$ then $e^{\ln x} \ln x = \ln y$ thus
$\ln  x =  W(\ln  y)$, leading  to  $x =  \exp(W(\ln  y)) =  \frac{\ln
  y}{W(\ln y)}$. Finally, we obtain
\[
\frac{n+c}{e} = \frac{\ln ((r/e)^{1/e})}{W(\ln ((r/e)^{1/e}))},
\]
and thus
\[
n = \frac{\ln(r/e)}{W(\frac{1}{e}\ln(r/e))} \cdot \left(1 +o(1) \right).
\]
The final estimation of $W$ is Eq.~(4.19) of \cite[page 349]{lambert}.
\end{proof}
The two  previous lemmas together with the  expression (\ref{as}) have
the following application:
\begin{pr}
Let  $p$ be  a prime  number, $n=O(p^{1/2})$,  $b_1,\ldots,b_{n-1}$ be
independently  and uniformly  distributed elements  in $\F_p$  and let
$f(x)  =  x^n +  \sum_{j=1}^{n-1}  b_{n-j}  x^j$.   When $n  \geqslant
\frac{\ln p }{\ln\ln  p} (1+o(1))$, the expected number  of element $k
\in \F_p$ such  that $f(x)-k$ splits into linear  factors in $\F_p$ is
$O(1)$.
\end{pr}
\begin{proof}
With    the   assumption    $n   =    O(p^{1/2})$,   the    value   of
$\binom{p+n-1}{n}/p^{n-1}$   is   a    constant   factor   away   from
$p/n!$. Solving the equation $n! = p$ via Lemma \ref{lambert} leads to
\[
n = \frac{\ln(p/e)}{\ln(1/e \ln(p/e))} \cdot (1+o(1)) =
\frac{\ln(p)}{\ln(\ln(p))} \cdot (1+o(1)).
\] 
Therefore as soon as the condition $n \geqslant \frac{\ln p }{\ln\ln p}
(1+o(1))$ is fulfilled, the conclusion of the proposition holds, due to
Lemma \ref{l1} and Eq.~(\ref{as}).
\end{proof}
The effective value $O(1)$ in  the above proposition is trivially 0 if
no such $k$ exists. Computer simulations tend to show that when $p$ is
reasonably large and  such a $k$ exists, as soon  as $n \geqslant \ln
p$, the value $O(1)$ is  1 with overwhelming probability, i.e., $k$ is
then unique.   Taking back the example described  in Section \ref{Wu},
when $p$  is a 75 bit  prime number, then as  soon as a  secret key is
broadcasted  to $n>15$ users,  being able  to solve  Eq.~(\ref{eq}) is
enough to recover $k$ with high probability.\\

The  consequence of  the above  proposition can  be summarized  in the
following terms. Any algorithm that  solves the problem of finding all
$k \in  \F_p$ such that $x^p-x =  0 \mod (f(x)-k)$, where  $f(x)$ is a
monic  $n$-th degree random  polynomial and  $n \geqslant  \frac{\ln p
}{\ln\ln p} \cdot (1+o(1))$, can be used to break the key distribution
scheme  in secure  multicasting of  Wu et  al. \cite{wu}  described in
Section \ref{Wu}.

%%%%%%%%%%%%%%%%%%%%%%%%%%%%%%%%%%%%%%%%%%%%%%%%%%%%%%%%%%%%%%%%%%%
\Section{Cryptanalysis of the Scheme}\label{At}

In this  section we present  two different approaches that  tackle the
security  of  the system.  The  first one  is  effective  when $n$  is
unusually large compared  to $p$, i.e., when $n$ is  not far away from
$p^{1/2}$. The second one uses  the existence of average size divisors
of $p-1$.
  
\subsection{Attack when $n=p^{1/2-\varepsilon}$ with
  small $\varepsilon$}

When  the number  of users  $n$  is large  compared to  $p$, a  simple
algebraic procedure can reveal with sufficiently large probability the
secret key $k$. The key point  is that the polynomial $f(x)$ takes the
value $k$  much more  often than  a random polynomial.  In fact  for a
truly random monic $n$-th degree polynomial $g$ the expected number of
roots of  $g(x)=k$ is one.  In  our case, it  is $n$. So for  a random
field  element $a$,  the probability  that  $f(a)=k$ is  $n/p$ and  by
computing
\[
r_a(x):= x^p-x \mod (f(x)-f(a)),
\]
we expect  to find $r_a(x)=0$ after  $p/n$ trials. In  view of Section
\ref{cy},  as soon  as  $n  \geqslant \frac{\ln  p  }{\ln\ln p}  \cdot
(1+o(1))$,   then   $a=h(a_i)$   for   some  $i$   with   overwhelming
probability. If the quotient $n/p$ is too small, then there is no hope
this   approach  can   lead  to   an  efficient   algorithm,   but  if
$n=p^{1/2-\varepsilon}$ with a small $\varepsilon$, then the situation
is  different.    Computing  $r_a$  requires   $O(\log_2  p)$  modular
polynomial  operations, which  leads  to an  attack  with an  expected
complexity  of  $O(p^{1/2+\varepsilon}   \ln  p)$  modular  polynomial
operations. For  example, when $p$  is a 64  bit number and $n$  is as
large as a million, i.e.,  $n\cong 2^{20}$, then $\varepsilon = 3/16$,
and  the  complexity  of   the  attack  is  roughly  $2^{50}$  modular
polynomial  operations,  compared  with  $2^{70}$ for  the  exhaustive
search on $k$ described in Section \ref{Wu}.

\subsection{Time-memory tradeoff attack}\label{TMT}

A more direct  approach to the problem of finding  an element $k$ such
that the modular  equation $x^p-x = 0 \mod (f(x)-  k)$ is fulfilled is
to consider $k$  as a variable and develop and  reduce the equation in
terms  of the  powers of  $k$.  More precisely,  since $f(x)  = x^n  +
\sum_{j=0}^{n-1} b_{n-j}  x^j$, then $x^n  = -\sum_{j=0}^{n-1} b_{n-j}
x^j +  k \mod (f(x)-  k)$, and the  power $x^p$ can be  reduced modulo
this  equality. In  other words,  when working  in $\F_p[x,y]$  we can
write
\[
x^p-x = \sum_{i=0}^{n-1} c_i(y) x^i \mod \left(f(x)-y \right).
\]
The polynomials  $c_i$ fulfill then  the condition $c_i(k)=0$  for all
$i$  since when  $y$ takes  the value  $k$, the  polynomial in  $x$ is
identically 0.  If we could  compute explicitly the  polynomials $c_i$
then we could  recover $k$ since with very  high probability $k$ would
be their only common root, and therefore
\[
x-k = \gcd\{c_i(x), \, i=0,\ldots,n-1\}.
\]
In any  case, the  number of linear  factors is  $O(1)$ as soon  as $n
\geqslant \frac{\ln  p }{\ln\ln p} (1+o(1))$, as  discussed in Section
\ref{cy} above.  However  one readily verifies that the  degree of the
$c_i$'s is $\lfloor p/n \rfloor$ and  in our case the memory needed to
work with these polynomials is unrealistic because $p/n$ is too large,
specially when  $n\ll p$. There exists  however a turn  around. Let us
factorize  the order  of $\F_p^*$  as $p-1=d_1d_2$  with  $d_1>1$.  If
$k \neq 0$ then $k^{d_1d_2}=k^{p-1}=1$ and thus $k^{d_1}$ can only take
$d_2$ values,  i.e., the $d_2$ roots  of unity in $\F_p$.   In fact if
$\beta$ is a primitive element of $\F_p$ and
\[
S:=\{\omega \in \F_p \; | \; \omega^{d_2}=1\} = \{\omega_j \; | \;
\omega_j = \beta^{j\frac{p-1}{d_2}} \mbox{ for some } j=0,\ldots,d_2-1\} 
\]
then  $k^{d_1}=\omega_j$ for some  $\omega_j \in  S$. Notice  that the
elements of $S$ can be  efficiently computed since primitive roots are
easily found, see e.g. \cite{gathen}.  For a given $\omega \in S$, let
$I_{\omega}$ be the ideal  in $\F_p[x,y]$ generated by the polynomials
$f(x)-y$ and $y^{d_1}-\omega$. In the quotient ring, we have
\[
x^p-x = \sum_{i=0}^{n-1} c_{i,\omega}(y) x^i \mod I_{\omega},
\]
where the polynomials $c_{i,\omega}$ satisfy $c_{i,\omega}(y) = c_i(y)
\mod (y^{d_1}-\omega)$.  Therefore,  the degrees of all $c_{i,\omega}$
are  bounded  by  $d_1-1$  and  when  $\omega  =  \omega_j$,  we  have
$c_{i,\omega}(k)=0$. The computation of the polynomials $c_{i,\omega}$
can be performed quite simply: when computing $x^p \mod I_{\omega}$ by
any square-and-multiply technique in  $\F_p[x,y]$, reduce at each step
all the terms of  degree larger or equal than $n$ for  $x$ with $x^n =
-\sum_{j=0}^{n-1}  b_{n-j} x^j  + y$  and those  larger or  equal than
$d_1$  for  $y$   with  $y^{d_1}=\omega$.   The  time-memory  tradeoff
algorithm  consists in  testing all  $d_2$ possible  $\omega$  until a
common linear  factor of the $n$ polynomials  $c_{i,\omega}$ is found,
revealing  the secret key  $k$.  Note  that the  cost of  the greatest
common  divisor   computation  is  $O(\ln   d_1)$  modular  polynomial
operations. The memory requirement is  $M=d_1 \log_2 p$ bits, the time
requirement is $T=O(d_2 \ln p \ln d_1)$ modular polynomial operations,
and we have $TM = O(p \ln^2 p\ln d_1)$.

Clearly the quality  of this approach depends on  the factorization of
$p-1$. The case where $p$ is a strong prime, see \cite{menezes}, i.e.,
$p=2q+1$,   with   $q$  prime,   is   immune   against  the   previous
attack. However as soon as $p-1$ has a factor $d_1$ with $t$ bits, and
if sufficient memory is available, then the time needed to compute the
secret key  from the public data  is decreased by a  factor of roughly
$2^t$  compared to  the brute  force described  earlier.  It  is worth
mentioning that the original scheme has no indication on the choice of
the special form of $p$. The  case of the example presented in Section
\ref{Wu} is illustrative.   When $p-1$, a 75 bit  number, has a factor
in  the range of  $2^{40}$, which  corresponds to  a few  gigabytes of
memory, the cost of the  attack is reduced to roughly $2^{45}$ modular
polynomial computations, much less than $2^{80}$, which corresponds to
the cost  of the brute  force search, and  is feasible by  an attacker
with realistic power.

Let us  briefly study the conditions  required in order  for the above
attack to terminate with a time and memory requirement in the order of
the square root of $p$. This boils down to determine how often a prime
$p$ is  such that $p-1$ has a  factor in the range  of $p^{1/2}$.  For
$0\leqslant   \alpha<\beta\leqslant    1$,   let   $N(x,   x^{\alpha},
x^{\beta})$ be  the number of primes  $p \leqslant x$  such that $p-1$
has  a  factor  $d$   such  that  $x^{\alpha}  \leqslant  d  \leqslant
x^{\beta}$.   There  exist  constants  $r$  and $B$,  that  depend  on
$\alpha$ and $\beta$, such that
\begin{equation}\label{h}
\forall x > B\;, \; \; N(x, x^{\alpha}, x^{\beta}) > \frac{rx}{\ln x}.
\end{equation}
This is  \cite[Theorem 7]{ford}.  Taking  into account that  there are
$\frac{x}{\ln  x}  (1+o(1))$ primes  smaller  than $x$,  Eq.~(\ref{h})
above states that for sufficiently large $x$, the proportion of primes
$p  \leqslant  x$ such  that  $p-1$ has  a  factor  in $[  x^{\alpha},
  x^{\beta} ]$  is larger  than a fixed  ratio. For  example, computer
simulations on prime  integers ranging from 30 bits  to 85 bits showed
that when $\alpha=0.475$ and  $\beta=0.5$, $r \geqslant 0.33$ seems to
fit  the reality. This  means that  for approximately  a third  of the
randomly chosen  large finite prime  fields, the above attacks  can be
mounted with a  time and memory complexity in the  range of the square
root of the field size.  The ratio jumps to $r>0.59$ for $\alpha=0.33$
and $\beta=0.5$,  corresponding to a time-memory tradeoff  of at least
$2/3$-$1/3$ bit complexity in almost 60 $\%$ of the cases.

%%%%%%%%%%%%%%%%%%%%%%%%%%%%%%%%%%%%%%%%%%%%%%%%%%%%%%%%%%%%%%%%%%%
\Section{Conclusion and Acknowledgments}

The key distribution system developed by Wu et al. aims at solving the
problem  of  key management  in  a  potentially insecure  multicasting
environment. We presented an analysis  of the system by shedding light
on the security  implied in the choices of the  two main parameters of
the scheme  $p$ and  $n$. Two different  attacks have  been presented,
both being efficient when  some conditions are fulfilled, exhibiting a
family of  weak parameters. For instance,  when $n\ll p$ and  $p$ is a
strong prime, the scheme is immune against both the attacks.

The author  would like to thanks  Jens Zumbr\"agel for  early talks on
this  subject,  as well  as  the people  of  the  Vienna Workshop  for
fruitful discussions.

% The authors  would like  to thank the  reviewer for  several helpful
% suggestions.

%%%%%%%%%%%%%%%%%%%%%%%%%%%%%%%%%%%%%%%%%%%%%%%%%%%%%%%%%%%%%%%%%%%


\begin{thebibliography}{10}

\bibitem{bettale}
        L. Bettale, J.-C. Faug\`ere and L. Perret.
        Hybrid approach for solving multivariate systems over finite fields.
        {\it Journal of Mathematical Cryptology},
        3 (2009), pp. 177--197.

\bibitem{lambert}
        R.M. Corless, G.H. Gonnet, D.E.G. Hare, D.J. Jeffrey and
        D. E. Knuth. 
        On the Lambert W function.
        {\it Advances in Computational Mathematics},
        Vol. 5 (1996), pp. 329--359.

\bibitem{faugere}
        J.-C. Faug\`ere and S. Rahmany.
        Solving systems of polynomial equations with symmetries using 
        SAGBI-Gr\"obner bases. 
        {\it Proceedings of the 2009 International Symposium on Symbolic 
          and Algebraic Computation},
        ACM, 2009, pp. 151--158. 

\bibitem{ford}
        K. Ford.
        The distribution of integers with divisors in a given
        interval.
        {\it Annals of Mathematics}, 
        Vol. 168 (2008), pp. 367--433.

\bibitem{gathen}
        J. von zur Gathen and J. Gerhard. 
        {\it Modern Computer Algebra}.
        2nd edition, Cambridge University Press, 2003.

\bibitem{mc1}
        C. Hadjicostis and  Y. Wu.
        On solving composite power polynomial equations. 
        {\it Math. Comput.},
        Vol. 74 , No. 250 (2005), pp. 853--868. 

\bibitem{lidle}
        R. Lidl and H. Niederreiter.
        {\it Finite Fields}.
        2nd edition, Cambridge University Press, 1997.

\bibitem{mead}
         D.G. Mead. 
         Newton's Identities. 
         {\it The American Mathematical Monthly}, 
         Vol. 99, No. 8 (1992), pp. 749--751.

\bibitem{menezes}
        A.J. Menezes, P.C. van Oorschot and S.A. Vanston.
        {\it Handbook of Applied Cryptography}.
        CRC Press, 2001.

\bibitem{wiki} 
        Cable television in the United States.
        {\it Wikipedia, The Free Encyclopedia}.
        Wikimedia Foundation, Inc.
        24 March 2012. Web. 8 April 2012.
        https:$//$en.wikipedia.org$/$wiki$/$Cable$\_$television$\_$in$\_$the$\_$United$\_$States$\#$Premium$\_$cable

\bibitem{wu}
        K.P. Wu, S.J. Ruan, F. Lai and C.K.Tseng.
        On key distribution in secure multicasting.
        {\it Proceedings of the 25th Annual IEEE Conference on Local
          Computer Networks}, 2000, pp. 208--212.

\bibitem{mc2}
        Y. Wu. 
        More on solving systems of power equations. 
        {\it Math. Comput.},
        Vol. 79, No. 272 (2010), pp. 2317--2332. 

\bibitem{zhu}
        W. T. Zhu.
        Cryptanalysis of Two Group Key Management Protocols for Secure Multicast.
        {\it CANS 2005}, Y.G. Desmedt et al. (Ed.),
        LNCS 3810, 
        2005, pp. 35--48.
        

\end{thebibliography}
\end{document}